\documentclass[aps,twocolumn,a4paper,showpacs,superscriptaddress]{revtex4}
\usepackage{graphicx}
\usepackage[all]{xy}
\usepackage{amsmath}
\usepackage{amssymb}
\usepackage{epstopdf}
\usepackage{dcolumn}
\usepackage{amsmath}
\usepackage{float}
\newcommand{\be}{\begin{equation}}
\newcommand{\ee}{\end{equation}}
\newcommand{\ben}{\begin{eqnarray}}
\newcommand{\een}{\end{eqnarray}}
\newcommand{\bes}{\begin{subequations}}
\newcommand{\ees}{\end{subequations}}

\begin{document}

\title{Compactlike kinks and vortices in generalized models}
\author{D. Bazeia}
\affiliation{Departamento de F\'{\i}sica, Universidade Federal da Para\'{\i}ba, 58051-970
Jo\~ao Pessoa, PB, Brazil}
\author{E. da Hora}
\affiliation{Departamento de F\'{\i}sica, Universidade Federal da Para\'{\i}ba, 58051-970
Jo\~ao Pessoa, PB, Brazil}
\author{R. Menezes}
\affiliation{Departamento de Ci\^encias Exatas, Universidade Federal da Para\'{\i}ba,
58297-000 Rio Tinto, PB, Brazil}
\affiliation{Centro de F\'\i sica e Departamento de F\'\i sica, Universidade do Porto, 4169-007 Porto, Portugal}
\author{H. P. de Oliveira}
\affiliation{Instituto de F\'{\i}sica, Universidade do Estado do Rio de Janeiro, 20513-013 Rio de Janeiro, RJ, Brazil}
\author{C. dos Santos}
\affiliation{Centro de F\'\i sica e Departamento de F\'\i sica, Universidade do Porto, 4169-007 Porto, Portugal}
\affiliation{Departamento de F\'\i sica, Universidade de Santiago de Compostela, 15782 Santiago de Compostela, Spain}


\begin{abstract}
{This work deals with the presence of topological defects in k-field models,
where the dynamics is generalized to include higher order power in the
kinetic term. We investigate kinks in (1,1) dimensions and vortices in (2,1)
dimensions, focusing on some specific features of the solutions. In
particular, we show how the kinks and vortices change to compactlike solutions,
controlled by the parameter used to introduce the generalized models.}
\end{abstract}

\maketitle

\section{Introduction}

In this work, we deal with defect solutions in k-field models, which are
models where the kinematics is generalized to allow for the presence of
terms depending on higher order power of the derivative of the fields.

There are several distinct motivations to study defect structures in high
energy physics. In the case of standard models, interest to investigate
defect structures can be found, for instance, in \cite{vs,ms}. Motivations
to study defect structures in generalized models come from Cosmology, with
the k-essence models \cite{ke1,ke2,ke3,ke4,ke5,ke6} and from other areas, as one can
find in the recent works \cite{k1,k2,k3,k4,k5,k6,k7,k8,k9,k10}.

A nice property of generalized models is that under specific conditions,
they may support compactons \cite{r}, which are defect solutions which live
in a compact region, so they have finite wavelength. This is different from
the standard defects, which are described by solutions of infinite
wavelength. Since compactons have gained recent interest in high
energy physics \cite{r2}, the main purpose of the present work is to
investigate the behavior of the defect solutions, and their modification
into compactons, under the variation of the driving parameter, which
responds for the generalized model. As we show below, the generalized models
which we will consider are controlled by a single real parameter, labeled $%
\alpha$, which responds for the generalization, in the sense that the limit $%
\alpha\to0$ leads us back to the standard model. This driving parameter $%
\alpha$ is then used to make the generalized model close to (for $\alpha$
small) or far away from (for $1/\alpha$ small) the standard model.

As one knows, the existence of excitations localized under the presence
of nonlinear interaction has long been explored with the hope to better understand the fundamental
contents of matter \cite{vs,ms}. In the case of compactons \cite{r,r2}, which are excitations characterized
by having a compact support, one notes that two adjacent compactons do not interact unless they come into
close contact. This is specific to compactons and in this sense, compact excitations seem to be well
appropriate to introduce new features as particlelike structures, as kinks in the line or vortices in the plane,
or immersed in space as domain walls or cosmic strings, respectively \cite{vs,ms,ke1,ke2,ke3,ke4,ke5,ke6,k1,k2,k3,k4,k5,k6,k7,k8,k9,k10}. 

In the present work we deal with kinks in generalized models described by a
single real scalar field $\phi$ in $(1,1)$ space-time dimensions. This is
done in the next Sec.~{\ref{sec:kinks}}, where we start with the standard
model and then generalize it and study the presence of defect structures.
Since the case of kinks is simpler, we use it to set the focus of the work,
to prepare for the study of vortices, which is done in Sec.~{\ref%
{sec:vortices}}. We deal with vortices considering generalized models in $%
(2,1)$ space-time dimensions, described by a complex scalar field $\varphi$
coupled to the $U(1)$ gauge field $A_\mu$, with the standard model being the
Maxwell-Higgs model firstly investigated in \cite{no}, with the
generalization being controlled by the parameter $\alpha$, in a way similar
to the case of kinks considered in Sec.~{\ref{sec:kinks}}. We end the work
in Sec.~{\ref{sec:end}}, where we introduce our comments and conclusions.

%%%%%%%%%%%%%%%%%%%%%%%%%%%%%%%%%%%%%%%%%%%%%%%%%%%

\section{The case of kinks}

\label{sec:kinks}

The model which describes a single real scalar field in $(1,1)$ space-time
dimensions is given by the Lagrange density 
\begin{equation}
{\mathcal{L}=\frac12\partial_\mu\phi\partial^\mu\phi-V(\phi)},
\end{equation}
or 
\begin{equation}
{\mathcal{L}=X-V(\phi)},
\end{equation}
where we have set 
\begin{equation}
{X=\frac12\partial_\mu\phi\partial^\mu\phi}.
\end{equation}

The equation of motion for static solution is 
\begin{equation}
\frac{d^2\phi}{dx^2}=\frac{dV}{d\phi}.
\end{equation}
It can be integrated to give 
\begin{equation}  \label{eqdeprimeiraordem}
\frac12\phi^{\prime2}=V(\phi).
\end{equation}
If we choose specific potential, we can obtain topological solution. As a
nice model, let us take the potential 
\begin{equation}  \label{phi4}
V(\phi)=\frac12(1-\phi^2)^2.
\end{equation}
In this case, we are using units such that both the field and coordinates
are dimensionless. Here we obtain the solution 
\begin{equation}  \label{kink}
\phi(x)=\tanh(x),
\end{equation}
after choosing its center to be at the origin $x=0$. The energy of the
solution is $E=4/3$. As we know, the width $w_k$ of the defect can be
written in terms of the height the potential between the two minima. Thus,
we can write 
\begin{equation}
w_k \propto {V^{-\frac12}(0)}.
\end{equation}
This shows that as higher the barrier between the two minima is, as thinner
the width of the defect solution is, when one fixes the distance between the
minima.

Let us now consider the generalized model 
\begin{equation}
\mathcal{L}=X-\alpha X^2-V(\phi),
\end{equation}
where $\alpha$ is a real parameter, introduced to control the modification
of the standard kinematics, with the limit $\alpha\to0$ leading us back to
the standard situation. Here the equation of motion for static solution is 
\begin{equation}
(1-3\alpha \phi^{\prime2} )\phi^{\prime\prime} = V_\phi.
\end{equation}
It can be integrated to give 
\begin{equation}
\frac12 \phi^{\prime2} + \frac{3\alpha}{4}\phi^{\prime4}=V(\phi),
\end{equation}
or 
\begin{equation}  \label{def}
\frac12 \phi^{\prime2}=U,
\end{equation}
where we have set 
\begin{equation}
U(\phi)=\frac16\,{\frac{\sqrt {1+12\,\alpha\,V(\phi)}-1}{\alpha}}.
\end{equation}
Note that the minima of the potential are given by $U(\phi)=0$, which
imposes that $V(\phi)=0$, as it happens to be the case in the standard
model. This shows that in the above description, the distance between the
minima does not depend on $\alpha$. However, for the potential given by Eq.~(%
\ref{phi4}), we get that 
\begin{equation}
U(0)=\frac16\,{\frac {\sqrt {1+6\,\alpha\,}-1}{\alpha}},
\end{equation}
and so the width of the defect in the generalized model depends on $\alpha$.
In fact, as we have investigated numerically, the dependence of the width on $\alpha$
is strong, and may make it difficult to understand how to get to the case of compacton
solutions. However, to make the study of the behavior of the defect structure for large $\alpha$
easier to understand, it is better to modify the potential $V(\phi)$ in a
way such that $U(0)$ is constant, independent of $\alpha$. Thus, we consider
the simple case in which $U(0)=1/2$. We get to this with the modification 
\begin{equation}
V(\phi)\to (1+\frac32\alpha)V(\phi),
\end{equation}
which leads to 
\begin{equation}
U(\phi)=\frac16\,{\frac{\sqrt {1+6\,\alpha\,(2+3\alpha)V(\phi)}-1}{\alpha}}.
\end{equation}

\begin{figure}[h!]
\begin{center}
\includegraphics[width=0.4\textwidth]{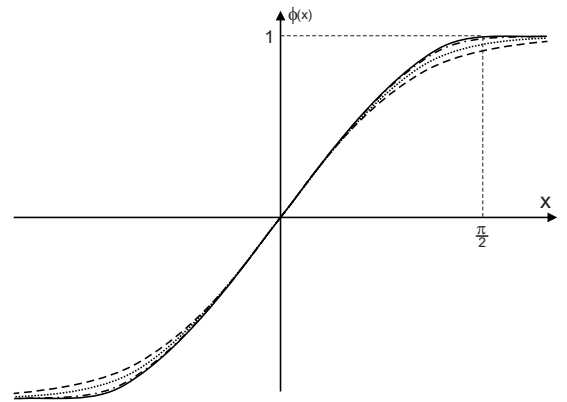}
\end{center}
\caption{The standard kink \eqref{kink} is shown with the dashed line, and
the compacton \eqref{comp} is shown with the solid line. The other solutions
are for other values of $\protect\alpha$, as we explain in Fig.~{\protect\ref{f2}}.}
\label{f1}
\end{figure}

With this at hand, it is now easy to understand the behavior of the solution for increasing
$\alpha$. We do this by first expanding the newer $U(\phi)$ in terms of $\alpha^{-1}$. In the case
of $\alpha\to\infty$ we get to 
\begin{equation}  \label{pcomp}
U(\phi)=\frac12 |1-\phi^2|\,.
\end{equation}
We use this into Eq.~\eqref{def} to get to the compacton 
\begin{equation}  \label{comp}
\phi(x)=%
\begin{cases}
1 & \mathrm{for}\,x>\frac{\pi}{2}, \\ 
\sin(x) & \mathrm{for}\,-\frac{\pi}{2}<x<\frac{\pi}{2}, \\ 
-1 & \mathrm{for}\,x<-\frac{\pi}{2}.
\end{cases}%
\end{equation}
In Fig.~{\ref{f1}} we plot both the kink \eqref{kink}, with dashed line, and
the compacton \eqref{comp}, with solid line, to show how the kink should
behave in the limit of very large $\alpha$. See also Fig.~2 for the other plots
shown in Fig.~1. 

We also depict in Fig.~3 the energy density $\rho(x)$ of the kink-like solution
for $\alpha=0,1$, and 10. We see from this figure that the energy density is localized,
irrespective of the value of alpha, as expected.

%%%%%%%%%%%%%%%%%%%%%%%%%%%%%%%%%%%%%%%%%%%%%%%%%%%%%%%%%%%%%%%%%%%
%%%%%%%%%%%%%%%%%%%%%%%%%%%%%%%%%%%%%%%%%%%%%%%%%%%%%%%%%%%%%%%%%%%%%%%%%%%%%

\begin{figure}[h!]
\begin{center}
\includegraphics[width=0.4\textwidth]{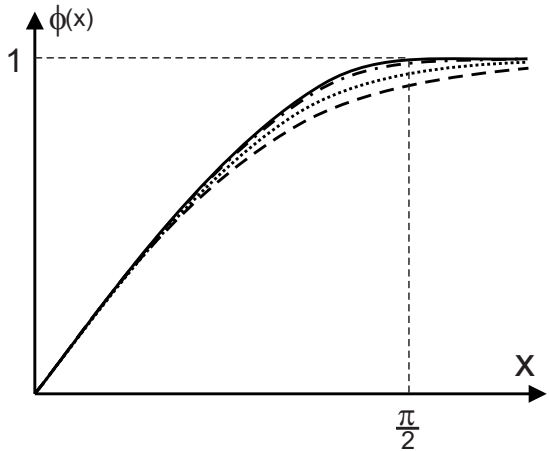}
\end{center}
\caption{A closer view of Fig.~1, showing the standard kink (dashed line), the compact solution
(solid line) and two other solutions, for $\alpha=1$ (dotted line) and for $\alpha=100$ (dash-dotted line).}
\label{f2}
\end{figure}

In the generalized model, the first-order equation which we have to solve
for $\alpha$ arbitrary is given by 
\begin{equation}  \label{defect}
\phi^{\prime2}=\frac13\,{\frac{\sqrt {1+3\,\alpha\,(2+3\alpha)(1-\phi^2)^2}-1}{\alpha}}.
\end{equation}
We numerically investigate this equation for several values of $\alpha$, and
we also plot some solutions in Fig~{\ref{f1}}: the dotted line is for $
\alpha=1$, and the dash-dotted line is for $\alpha=100$. The behavior of the
solution for varying $\alpha$ is better seem in Fig.~{\ref{f2}}. There we
see that for increasing $\alpha$ the defect converges to the compacton shown
in Fig.~\eqref{f1}, as expected. However, the convergence is very slow since
one needs a very large $\alpha$ to make the kink to behave as a compacton.

%%%%%%%%%%%%%%%%%%%%%%%%%%%%%%%%%%%%%%%%%%%%%%%%%%%%%%%%%
%%%%%%%%%%%%%%%%%%%%%%%%%%%%%%%%%%%%%%%%%%%%%%%%%%%%%%%%%%%%%

The above investigation help us to understand the basic behavior of the
defect solutions, and the passage to compactons. It will help us to
understand the much more complicated situation, where we deal with vortices,
to investigate the presence of compactlike vortices in generalized models. This
is the subject of the next Sec.~\ref{sec:vortices}.

%%%%%%%%%%%%%%%%%%%%%%%%%%%%%%%%%%%%%%%%%

\section{The case of vortices}
\label{sec:vortices}

Let us first introduce our model. It is described by the Lagrange density
\begin{equation}
\mathcal{L}=-\frac{1}{4}F_{\mu \nu }F^{\mu \nu }+\left\vert D_{\mu }\phi \right\vert^{2}-\alpha \left\vert D_{\mu}\phi\right\vert^{4}-V(|\phi|^2)\text{ ,}
\end{equation}
where $F_{\mu\nu}=\partial_\mu A_\nu-\partial_\nu A_\mu$ and 
\begin{equation}
D_{\mu}\phi=\partial_{\mu}\phi+ieA_{\mu}\phi,
\end{equation}
and $V(|\phi|^2)$ is the potential which implements spontaneous symmetry breaking.

As before, here $\alpha$ is a real parameter which controls the generalized dynamics, with $\alpha\to0$ leading us back to the standard Maxwell-Higgs model.
Also, $e$ stands for the electric charge, and the potential usually has two parameters, $\lambda$ and $v$, which represent the coupling constant for self-interaction of the scalar field and the spontaneous symmetry breaking parameter, respectively. Usually, it is given by
\be
V(\phi)=\frac{\lambda^2}{4}\left(v^2-|\phi|^2\right)^{2}.
\ee
However, we are working in $(2,1)$ space-time dimensions, and so the fields, parameters and coordinates are not dimensionless quantities. For simplicity, however, we can rescale fields, parameters and coordinates in order to work with dimensionless quantities. We do this in the usual
way, and we can write the new Lagrange density in the simpler form
\begin{equation}
\mathcal{L}=-\frac{1}{4}F_{\mu \nu }F^{\mu \nu }+\left\vert D_{\mu }\phi \right\vert^{2}-\alpha \left\vert D_{\mu}\phi\right\vert^{4}-V(|\phi|^2)\text{ ,}
\end{equation}
where we are now using dimensionless fields, coordinates and $\alpha$, with $\lambda=v=e$, for simplicity. We are also using the same $F_{\mu\nu}=\partial_\mu A_\nu-\partial_\nu A_\mu$, but now we have changed the covariant derivative to
\begin{equation}
D_{\mu}\phi=\partial_{\mu}\phi+iA_{\mu}\phi.
\end{equation}
In this case, we can consider the potential in the form
\begin{equation}
V(\phi)=\frac{1}{4}\left(1+\frac32\alpha\right)\left(1-|\phi|^2\right)^{2}.
\end{equation}
Note that the potential has the same form used in the case of kinks, and the presence of $\alpha$ has the same motivation there considered. Note also that we could have introduced
other models, one of them with another extra $\alpha$-dependent contributions added to the Maxwell term, for instance. However, we have implemented the above modification since it leads to the simplest model which can be obtained starting from the scalar field model used in the former Sec.~{\ref{sec:kinks}}.

%%%%%%%%%%%%%%%%%%%%%%%%%%%%%%%%%%%%%%%%%%%%%%%%%%%%%%%%%
\begin{figure}[t]
\begin{center}
\includegraphics[width=0.4\textwidth]{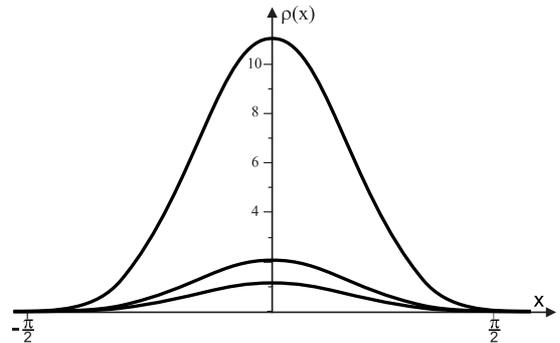}
\end{center}
\caption{The energy density $\rho(x)$ of the kink-like solution of Eq.~{\eqref{defect}} for $\alpha=0, 1$ and $10$, with greater $\alpha$ corresponding to higher solid curve.}
\label{f3a}
\end{figure}
%%%%%%%%%%%%%%%%%%%%%%%%%%%%%%%%%%%%%%%%%%%%%%%%%%%%%%%%%%%%%

In the investigation below, for simplicity we will sometimes use
\be
Y=|D_\mu\phi|^2
\ee 
and 
\be
F(Y)=Y-\alpha Y^2
\ee
to describe the generalization included in the above model in a shorter way.

The equations of motion are now given by
\begin{equation}
\partial _{\alpha }F^{\mu \alpha }=J^{\mu }\text{ ,}
\end{equation}
and
\begin{equation}
F_{{Y}}{D}^{\mu }{D}_{\mu }\phi +F_{{YY}}{D}^{\mu }\phi \partial_\mu{Y}=-\frac{\partial V}{\partial 
\overline{\phi }}\text{ ,}
\end{equation}%
where
\begin{equation}
J^{\mu }=i\left(\phi \overline{{D}}^{\mu }\overline{\phi}-\overline{\phi }{D}^{\mu}\phi \right) F_{Y}\text{ ,}
\end{equation}%
with $F_{{Y}}={dF}/{d{Y}}$ and $F_{{YY}}={d^{2}F}/{d{Y}^{2}}$.

The Gauss Law is written as
\begin{equation}
\nabla^{2}A^{0}=2 A^{0}\left\vert \phi
\right\vert ^{2}F_{{Y}}\text{ ,}
\end{equation}
which allows that we use $A^{0}=0$ as a proper gauge choice. We
fix this temporal gauge from now on.

In order to investigate the presence of vortices, let us consider the standard static and rotationally symmetric \textit{Ansatz}, which implies that
\begin{equation}\label{Ans1}
\phi (r,\theta )=g\left( r\right) \exp \left( in\theta \right),
\end{equation}%
\begin{equation}\label{Ans2}
\mathbf{A}(r,\theta )=-\frac{1}{r}\left(a(r)-n\right) 
\hat{\theta },
\end{equation}
where $n=\pm1,\pm2,...$ describes the vorticity of the solution. In this case, the dimensionless static and rotationally symmetric equations of motion become
\be\label{em1}
\frac{d^{2}a}{dr^{2}}-\frac{1}{r}\frac{da}{dr}-4\alpha K_1(g,a)=2g^{2}a\text{ ,}
\ee
and
\be\label{em2}
\frac{d^{2}g}{dr^{2}}+\frac{1}{r}\frac{dg}{dr}-\frac{a^{2}g}{r^{2}}+2\alpha K_2(g,a)=\left(1+\frac32 \alpha\right)(g^3-g).
\ee
where $K_1(g,a)$ and $K_2(g,a)$ are given by
\be
K_1(g,a)=g^{2}a\left( \frac{dg}{dr}\right)^{2}+\frac{g^{4}a^{3}}{r^{2}},
\ee
and
\ben
K_2(g,a)&=&\frac{1}{r}\left(\frac{dg}{dr}\right)^{3}+3\left(\frac{dg}{dr}\right)^{2}\frac{d^{2}g}{dr^{2}}+\frac{g^{2}a^{2}}{r^{2}}\frac{d^{2}g}{dr^{2}}+\nonumber\\
&&\frac{ga}{r^{2}}\frac{dg}{dr}\left({a}\frac{dg}{dr}+2g\frac{da}{dr}-\frac{ga}{r}\right)-\frac{g^{3}a^{4}}{r^{4}}.
\een

In the limit $\alpha\to0$, the above equations get to
\be
\frac{d^{2}a}{dr^{2}}-\frac{1}{r}\frac{da}{dr}=2g^{2}a\text{ ,}
\ee
and
\be
\frac{d^{2}g}{dr^{2}}+\frac{1}{r}\frac{dg}{dr}-\frac{a^{2}g}{r^{2}}=g^3-g.
\ee
which exactly reproduces the equations of motion of the standard Maxwell-Higgs model. According to our conventions, here we are dealing with scalar and vector fields with the same mass. Thus,
we can write the first order equations
\be
\frac{dg}{dr}=\pm \frac1r g\,a
\ee
and
\be
\frac1r\,\frac{da}{dr}=\mp (1-g^2)
\ee
Their solutions are BPS states, since they solve the equations of motion and have energy minimized to the Bogomol'nyi bound. This case is well-understood and can be found, for instance, in Ref.~\cite{vs}.

The generalized model is much more complicated. To understand the main features for a non-vanishing $\alpha$, let us consider the energy-momentum tensor. It has the form 
\ben
T_{\mu \nu }&=&\frac{2}{\sqrt{\left(-\eta\right)}}\frac{\partial\left[\sqrt{\left( -\eta \right) }\mathcal{L}\right] }{\partial \eta ^{\mu \nu }}\nonumber\\
&=&-\eta _{\mu \nu }\mathcal{L}-F_{\mu \alpha }F_{\nu }{}^{\alpha }+2F_{{Y}}{D}_{\nu }\phi\overline{{D}}_{\mu }\overline{%
\phi }\text{ ,}
\een
where $\eta ^{\mu \nu }=(+--)$ identifies the metric signature. Thus, the energy density is given by
\ben
T_{00}=\rho &=&\frac{{B}^{2}}{2}-{Y}+\alpha {Y}^{2}+\frac{1}{4}\left(1+\frac32 \alpha\right)\left( 1-|\phi|^2\right)^2\nonumber\\
&=&\frac{{B}^{2}}{2}+\left\vert \mathbf{D}\phi \right\vert^{2}+\alpha\left\vert\mathbf{D}\phi\right\vert^{4}+\nonumber\\
&&\frac{1}{4}\left(1+\frac32 \alpha\right)\left(1-|\phi|^2\right)^2\text{ ,}
\een
where $B$ is the magnetic field and 
\begin{equation}
{Y}=-\left\vert \mathbf{D}\phi \right\vert ^{2}=-\left\vert 
\overrightarrow{\nabla }\phi -i\mathbf{A}\phi \right\vert
^{2}=-\left( \frac{dg}{dr}\right) ^{2}-\frac{g^{2}a^{2}}{r^{2}}
\text{ .}
\end{equation}
The total energy of the static solution has the form
\ben
E&=&\int \rho(r)\; d^{2}r\nonumber\\
&=&2\pi\int rdr\biggl(\frac{{B}^{2}}{2}\!+\!\left\vert \mathbf{D}\phi
\right\vert^{2}\!+\!\alpha\left\vert \mathbf{D}\phi \right\vert^{4}+\nonumber\\
&&\;\;\;\;\;\;\;\;\frac{1}{4}\;\Bigl(1+\frac32 \alpha\Bigr)\left( 1- |\phi|^2\right)^{2}\biggr),
\een
and the presence of $\alpha$ destroys the Bogomol'nyi bound. We then come to the conclusion that for the generalized model at hand we have to study the equations of motion, and this requires numerical investigation.

To solve the equations of motion with finite energy solutions, the boundary conditions on $g(r)$ and $a(r)$ are given by,
near the origin
\be
\underset{r\rightarrow 0}{\lim }g\left( r\right) \rightarrow0\text{\;\;\; and \;\;\;}\underset{r\rightarrow 0}{\lim }a\left( r\right) \rightarrow n\text{,} \label{bc1}
\ee
and at very large distances
\be
\underset{r\rightarrow \infty }{\lim }g\left( r\right) \rightarrow 1 \text{\;\;\; and \;\;\; }\underset{r\rightarrow \infty }{\lim }a\left( r\right) \rightarrow 0 \label{bc2}
\text{.}
\ee

We have investigated the equations of motion \eqref{em1} and \eqref{em2} numerically, and the results for $a(r)$ and $g(r)$ are depicted in Figs.~4 and 5 for some values of $\alpha$.

%%%%%%%%%%%%%%%%%%%%%%%%%%%%%%%%%%%%%%%%%%%%%%%%%%%%%%%%%%%%%%%
\begin{figure}[t]
\begin{center}
\includegraphics[width=0.35\textwidth]{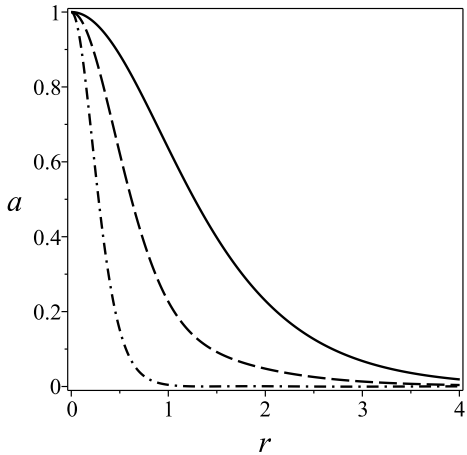}
\end{center}
\caption{The numerical solution of the equation of motion for $a(r)$ for $\alpha=0,10$ and 100, shown with the solid,
dashed and dash-dotted lines, respectively.}
\end{figure}
%%%%%%%%%%%%%%%%%%%%%%%%%%%%%%%%%%%%%%%%%%%%%%%%%%%%%%%%%%%%%%

The numerical strategy was to use the pseudospectral method \cite{pseudo} that consisted in approximating the fields $a(r)$ and $g(r)$ by
\ben
a(r) \simeq a_N(r) = \sum_{k=0}^N\,\hat{a}_k\psi_k(r), \\
g(r) \simeq g_N(r) = \sum_{k=0}^N\,\hat{g}_k\chi_k(r),
\een
where $N$ is the truncation order that dictates the number of modes $\hat{a}_k$ and $\hat{g}_k$ kept in the series, with $\psi_k(r)$ and $\chi_k(r)$ being the basis functions expressed in terms of the Chebyshev polynomials, and reproducing the boundary conditions (\ref{bc1}) and (\ref{bc2}). 

%%%%%%%%%%%%%%%%%%%%%%%%%%%%%%%%%%%%%%%%%%%%%%%%%%%%%%%%%%%%%%%
\begin{figure}[t!]
\begin{center}
\includegraphics[width=0.35\textwidth]{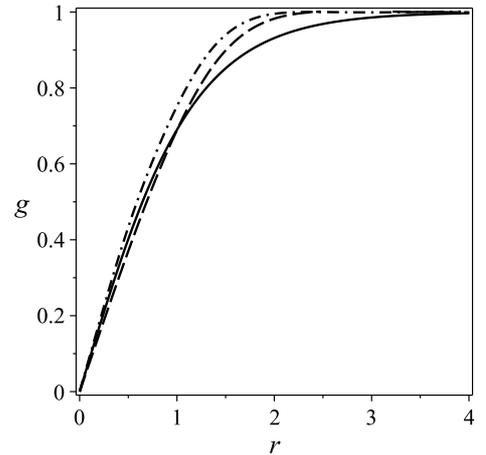}
\end{center}
\caption{The numerical solution of the equation of motion for $g(r)$ for $\alpha=0,10$ and 100, shown with the solid,
dashed and dash-dotted lines, respectively.}
\end{figure}
%%%%%%%%%%%%%%%%%%%%%%%%%%%%%%%%%%%%%%%%%%%%%%%%%%%%%%%%%%%%%%

We have used $N=15$ and followed straightforwardly the steps for the implementation of the pseudospectral method in the present case resulting in the determination of the modes $\hat{a}_k$ and $\hat{g}_k$ yielding, as a consequence, the reconstruction of the fields $a(r)$ and $g(r)$. The numerical results are obtained for $n=1$. In Figs.~4 and 5 we depict the fields $a(r)$ and $g(r)$ for several values of $\alpha$. We note from these figures that both $a(r)$ and $g(r)$ go to the corresponding vacuum states very rapidly, for increasing values of the parameter $\alpha$. The results show that for larges values of $\alpha$, the vortices behave as compactlike solutions, becoming constant field configurations at some finite distance from the origin. 

An important way to study topological structures requires that we search for the conserved topological charge, which in the case of vortices is the flux of the magnetic field. 
To see this, let us introduce the topological current
\be
J^\mu=\epsilon^{\mu\nu\lambda}\partial_\nu A_\lambda
\ee
which is conserved. Thus, we can write the topological change density
\be
J^0=\frac{\partial A_y}{\partial x}-\frac{\partial A_x}{\partial y}.
\ee
We can use the {\it Ansatz} for $\bf A$ given by Eq.~{\eqref{Ans2}} to see that $J^0=B$, and the topological charge density equals the magnetic field $B$.
Thus, the topological charge has the form
\be
Q_T=\int d^2r B=\Phi_B,
\ee 
and it gives the flux $\Phi_B$ of the magnetic field in the plane.

For this reason, let us then investigate the dimensionless magnetic field, which is given by 
\be
B=-\frac1r\frac{da}{dr}.
\ee
We use the boundary conditions on $a(r)$ to see that
\be
Q_T=\Phi_B=2\pi n,
\ee
showing that the topological charge or the flux of the magnetic field is conserved, and it is quantized according to the winding number $n=\pm1,\pm2,...$
  
In fig.~6 we plot the magnetic field for $\alpha=0,1,5$ and $10$, and there we see that it has the appropriate feature, showing the compactlike behavior of the vortex as $\alpha$ increases.
This is a nice behavior, because the magnetic field is gauge invariant, and it is directly related to the topological charge of the planar vortices.
We also plot in Fig.~7 the energy density of the vortices for $\alpha=0,1,5$, and $10$. We note that the energy density does not show the same behavior of the magnetic field, because it is localized irrespective of the value of $\alpha$, as expected.

%%%%%%%%%%%%%%%%%%%%%%%%%%%%%%%%%%%%%%%%%%%%%%%%%%%%%%%%%%%%%%
\begin{figure}[h!]
\begin{center}
\includegraphics[width=0.35\textwidth]{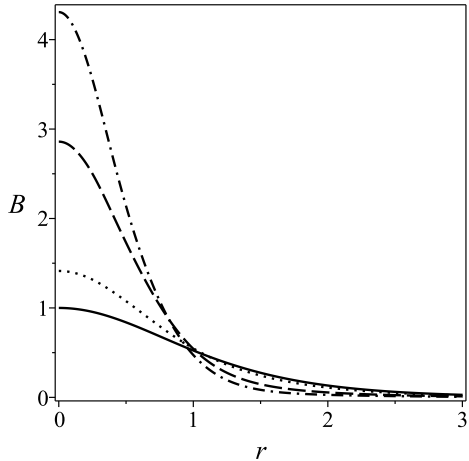}
\end{center}
\caption{The dimensionless magnetic field is plotted for $\alpha=0,1,5$ and 10, with the solid, dotted, dashed and dash-dotted lines, respectively.}
\end{figure}
%%%%%%%%%%%%%%%%%%%%%%%%%%%%%%%%%%%%%%%%%%%%%%%%%%%%%%%%%%%%%%%%%%%

%%%%%%%%%%%%%%%%%%%%%%%%%%%%%%%%%%%%%%%%%%%%%%%%%%%%%%
\section{Ending comments}
\label{sec:end}

In this work we have dealt with the presence of defect structures in $(1,1)$ and in $(2,1)$ space-time dimensions.
In $(1,1)$ space-time dimensions, we focused on kinks and the corresponding compactlike solutions in generalized models,
controlled by a single parameter $\alpha$. In this case, we have verified that the presence of $\alpha$ makes the
study somehow complicated, and so we have changed the potential, in order to simplify the investigation. The modification
introduced very much help us to clearly see the behavior of the solution for increasing $\alpha$, reaching compactlike
features for large values of $\alpha$. 

In $(2,1)$ space-time dimensions, we have investigated the presence of vortices in a generalized model, with the generalization
being also controlled by the single real parameter $\alpha$. The case of vortices is quite different from the case of kinks, but here 
we have solved the equations of motion for several values of $\alpha$, and we have identified the compactlike behavior in the vortex solution
which we found numerically. This seems to be a new behavior of the vortices, and we hope that the present investigation will stimulate new investigations
in the field, mainly on the main features the compactlike vortices may engender. In particular, we need more numerical investigations to see if the
compactlike behavior which we have found can lead to compact vortices. In this sense, an interesting issue could be to investigate if the compactlike vortices obey
the behavior found before for standard vortices \cite{wei}. Another issue concerns the natural extension of the present work to the case of generalized
model in the presence of the Chern-Simons dynamics \cite{cs1,cs2}.

%%%%%%%%%%%%%%%%%%%%%%%%%%%%%%%%%%%%%%%%%%%%%%%%%%%%%%%
\begin{figure}[h!]
\begin{center}
\includegraphics[width=0.35\textwidth]{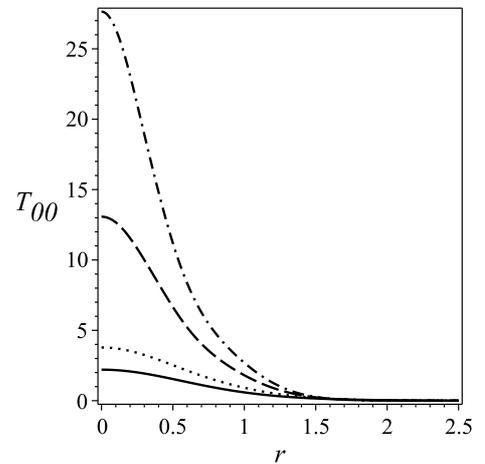}
\end{center}
\caption{Energy density of the vortices for $\alpha=0,1,5$ and 10, with the solid, dotted, dashed and dash-dotted lines, respectively.}
\end{figure}
%%%%%%%%%%%%%%%%%%%%%%%%%%%%%%%%%%%%%%%%%%%%%%%%%

The Brazilian authors would like to thank CAPES, CNPq and FAPERJ for partial financial support.
C. dS. is supported by the FCT Grants SFRH/BSAB/925/2009 and CERN/FP/109306/2009.

%%%%%%%%%%%%%%%%%%%%%%%%%%%%%%%%%%%%%%%

\end{document}